\documentclass[showpacs,onecolumn,superscriptaddress,aps]{revtex4}%
\usepackage{epsfig,dsfont,amssymb,amsmath,amsthm,amsfonts,amsbsy,mathrsfs}
\usepackage{graphicx}
\usepackage{amsmath}
\usepackage{amssymb}
\usepackage{lipsum}
\usepackage{amsfonts}%
\begin{document}
\title{Highly Efficient Processing of Multi-photon States}
\author{Qing Lin}
\email{qlin@mail.ustc.edu.cn}
\affiliation{Fujian Provincial Key Laboratory of Light Propagation and Transformation, College of Information Science and Engineering, Huaqiao University,
Xiamen 361021, China}
\author{Bing He}
\email{binghe@uark.edu}
\affiliation{Department of Physics, University of Arkansas, Fayetteville, AR 72701, USA}


\begin{abstract}
How to implement multi-qubit gates is an important problem in
quantum information processing. Based on cross phase
modulation, we present an approach to realizing a family of multi-qubit gates 
that deterministically operate on single photons as the qubits.
A general $n$-qubit unitary operation is a typical example of these gates.
The approach greatly relax the requirement on the resources, such as the ancilla photons and coherent beams, 
as well as the number of operations on the qubits. 
The improvement in this framework may facilitate large scale quantum
information processing.

\end{abstract}
\maketitle

\section*{Introduction} 

Recently the research on quantum information processing has been approaching the stage 
of large number of qubits, and numerous exciting developments in this trend have been reported. 
In 2009, Monz \textit{et al.} reported the creation of 14-qubit entanglement
with trapped ions \cite{trapion}. Later in 2011 and 2012, Huang \textit{et al.}
\cite{Guo} and Yao \textit{et al.} \cite{Pan} reported the generation of eight-photon
entanglement, respectively. How to efficiently process the ever increasing number of qubits becomes a 
prominent problem. Usually people follows the practice of classical computation to decompose a circuit 
into CNOT gates as the basic two-qubit gate and single-qubit gates. This CNOT-based approach
has been adopted in many research works, and its target is to decompose
a quantum circuit into as few CNOT gates as possible \cite{Knill, Barenco,
Smolin, Cybenko, Aho, Zhang, Vidal, Mottonen, Shende1, Vartiainen, Shende2}.
For a two-qubit unitary operation, at least three CNOT gates should be required
\cite{Zhang, Vidal}; for the well-known $n$-qubit Toffoli gates,
$\mathcal{O}\left(  n^{2}\right)  $ CNOT gates are necessary \cite{Barenco}.
Meanwhile, $2^{n-1}$ CNOT gates should be used to implement a general $(n-1)$-control-$1$ 
gate \cite{Mottonen}. The theoretical lower bound for a general $n$-qubit unitary
operation is $\mathcal{O}\left(  4^{n-1}\right)  $ CNOT gates \cite{Shende1},
but the actual number should be up to $\frac{23}{48}4^{n}$ \cite{Shende2}.

As early as in 2001, Knill, Laflamme and Milburn proposed the architecture 
of optical quantum computation, based on the prepared multi-photon entangled states as the ancilla 
\cite{KLM}. In the same year, Pittman \textit{et al.} presented a scheme to realize CNOT gate based only on linear
optics \cite{Pittman}. The CNOT gate has been experimentally demonstrated with linear optics
\cite{CNOT1, CNOT2, CNOT3, CNOT4}, as well as another two-qubit photonic gate, c-phase
gate \cite{cphase1, cphase2}. Moreover, numerous multi-qubit photonic gates, e.g. the well known Fredkin and Toffoli gate, were proposed 
\cite{Fredkin1, Fredkin2, Fredkin3, Toffoli1} and implemented \cite{Toffoli2} with linear optics.

Though the linear optical realization of photonic quantum circuits is possible, due
to the probabilistic nature in operation, its efficiency will not be so high when the processed photon number becomes large. 
A significant improvement is using weak cross-Kerr nonlinearity, so that the photonic gate operations could be made deterministic. 
The first application of weak cross-phase modulation (XPM) is the parity check \cite{Barrett}. 
Afterwards, this technique was adopted to an implementation proposal of deterministic CNOT gate \cite{munro}. 
Following this line of research, many applications of weak Kerr nonlinearity in quantum information processing have been proposed 
in recent years \cite{mun, Spiller1, Spiller2, req1, req2, dis, Loock, rep, tr, Sheng0, Sheng1, c-state, Sheng2, guo, g-state, Sheng3, Deng, Sheng4, wang, lee, zhang, Sheng5, li-ghose}. Moreover, photon loss and
decoherence effects in weak Kerr nonlinearity were studied from different perspectives \cite{Jeong1, Jeong2, SDBarrett, Louis, real}. 
Going back to quantum computation, the circuit construction in most of previous works belongs
to the CNOT-based scenario, i.e. CNOT gate is the elementary gate for constructing any quantum circuit. An alternative route to quantum computation is based on a different type of basic logic gates, c-path and merging gate. The first design of c-path gate based on weak XPM was introduced in 2009 \cite{qlin1}, and later the design was simplified and developed to a deterministic one \cite{qlin2, qlin3, qlin4}. This c-path-merging approach has been experimentally demonstrated with linear optical elements 
\cite{Brien}, and the essential idea of c-path and merging gate has also been utilized in other experiments \cite{CE1, CE2, CE3}.
These two universal elementary logic gates can efficiently realize various controlled logic gates such
as Fredkin and Toffoli gate. Especially the construction with c-path and merging gates can reduce the complexity 
of a Toffoli gate from polynomial $\mathcal{O}(n^{2})$ to linear \cite{qlin2, qlin3}.

In this paper, based on the improved designs for c-path and merging gate, we will
develop the c-path-merging approach to realize various multi-qubit controlled 
unitary operations and the general $n$-qubit unitary operation. Compared with the
CNOT-based approach, various controlled unitary operations can be implemented
more efficiently with less resources and less operations. We will show that,
for the realization of a general $(n-1)$-control-$1$ unitary operation, the
required resources, e.g. ancilla coherent states, ancilla single photons, can be reduced from exponential 
to linear, providing an optimization of such unitary operation. Furthermore, two approaches for 
realizing a general $n$-qubit unitary operation are proposed.

The rest of the paper is organized as follows. We firstly present the
improved optical realization of the two element gates, c-path and merging gate.
These element gates will be used to construct various multi-control gates
and the general unitary operation in next part. Afterward, we will discuss 
the complexity of our approach and compare it with the CNOT-based approach. 
Then the paper is concluded with the final part.

\section*{Element logic gates}

The operations described below are performed by two element gates, c-path and merging
gate \cite{qlin1,qlin2,qlin3}. Here, for the purpose of clarity, we will first describe a special example 
of c-path gate and then develop it to more control and target modes. After that we will improve the original merging gate by dispensing with the ancilla
single photon in its original design, and also show that it can be generalized to more spatial modes. Compared with the
former works \cite{qlin1,qlin2,qlin3, qlin4}, the primary advantage in the current approach is that no ancilla photon is necessary 
to any circuit, no matter how complicated it could be.

\subsection*{C-path gate}

This element gate encodes the bit information of a control qubit into the spatial modes of the target qubit. 
In Fig.1 the realization of an example of c-path gate is shown. Here we adopt the definitions $\left\vert
0\right\rangle \equiv\left\vert H\right\rangle $ and $\left\vert
1\right\rangle \equiv\left\vert V\right\rangle $, where H and V represent 
the two polarizations of a single photon, respectively. The input state is as follows,%

\begin{align}
\left\vert \Psi\right\rangle  =\left\vert H\right\rangle _{C}\left\vert \phi_{1}\right\rangle _{T}  +\left\vert V\right\rangle _{C}\left\vert \phi
_{2}\right\rangle _{T}, \label{in1} 
\end{align}
where the states $\left\vert \phi_{1(2)}\right\rangle $ are in arbitrary
forms ($\alpha_{i}\left\vert H\right\rangle +\beta_{i}\left\vert
V\right\rangle $, with $\sum\limits_{i=1}^{2}\left(  \left\vert \alpha
_{i}\right\vert ^{2}+\left\vert \beta_{i}\right\vert ^{2}\right)  =1$. At
firstly, let the control photon transmit through a polarized beamsplitter (PBS)
and the target photon through a 50:50 beamsplitter (BS). The  
target photon will be separated into $2$ spatial modes ($1, 2$). Secondly, one introduces two
coherent states $\left\vert \alpha\right\rangle_{cs} \left\vert \alpha\right\rangle_{cs}
$ (qubus beams) and let them interact with the input two single photons
through XPM as shown in Fig. 1. Then the input state will evolve to the
following state
\begin{align}
&  \frac{1}{\sqrt{2}}\left\{  \left\vert H\right\rangle _{C}\left\vert \phi_{1}\right\rangle _{1} \left\vert
\alpha e^{i\theta}\right\rangle_{cs} \left\vert \alpha e^{i\theta}\right\rangle_{cs}
\right.  +\left\vert H\right\rangle _{C}\left\vert \phi_{1}\right\rangle _{2} \left\vert \alpha\right\rangle_{cs} \left\vert
\alpha e^{i2\theta}\right\rangle_{cs} \nonumber\\
&  +\left\vert V\right\rangle _{C}\left\vert \phi
_{2}\right\rangle _{1}\left\vert \alpha e^{i2\theta
}\right\rangle_{cs} \left\vert \alpha\right\rangle_{cs}   \left.  +\left\vert V\right\rangle _{C}\left\vert \phi
_{2}\right\rangle _{2}\left\vert \alpha e^{i\theta
}\right\rangle_{cs} \left\vert \alpha e^{i\theta}\right\rangle_{cs} \right\}  .
\end{align}
Here we assume the XPM between single-mode coherent state and
single-mode one photon state, which is valid under the conditions specified in
\cite{xpm-01, xpm-02}. After that, a phase shifter of $-\theta$ is,
respectively, applied to two qubus beams, followed by the transformation $\left\vert \alpha_{1}\right\rangle_{cs} \left\vert
\alpha_{2}\right\rangle_{cs} \rightarrow\left\vert \frac{\alpha_{1}-\alpha_{2}%
}{\sqrt{2}}\right\rangle_{cs} \left\vert \frac{\alpha_{1}+\alpha_{2}}{\sqrt{2}%
}\right\rangle_{cs} $ with one more 50:50 BS on the coherent state components. 
The state of the total system will be therefore transformed to%

\begin{align}
&  \frac{1}{\sqrt{2}}\left\{  \left\vert H\right\rangle _{C}\left\vert \phi_{1}\right\rangle _{1} \left\vert
0\right\rangle_{cs} \left\vert \sqrt{2}\alpha\right\rangle_{cs}
\right.  +\left\vert H\right\rangle _{C}\left\vert \phi_{1}\right\rangle _{2} \left\vert -\beta\right\rangle_{cs} \left\vert
\sqrt{2}\alpha\cos\theta\right\rangle_{cs} \nonumber\\
&  +\left\vert V\right\rangle _{C}\left\vert \phi
_{2}\right\rangle _{1}\left\vert \beta\right\rangle_{cs} \left\vert
\sqrt{2}\alpha\cos\theta\right\rangle_{cs}   \left.  +\left\vert V\right\rangle _{C}\left\vert \phi
_{2}\right\rangle _{2}\left\vert
0\right\rangle_{cs} \left\vert \sqrt{2}\alpha\right\rangle_{cs} \right\}  .
\end{align}
where $\left\vert \beta\right\rangle_{cs} =\left\vert i\sqrt{2}\alpha\sin
\theta\right\rangle_{cs}$ and $\left\vert0\right\rangle_{cs}$ denotes the coherent vacuum state. To obtain the desired state, we need a photon number-resolving
detector (PND) to measure the first qubus beam ($\left\vert \pm\beta\right\rangle_{cs}$), and then separate the first and fourth components from the second and third components in the above state. The photon number resolution, denoted by the projection $\left\vert k\right\rangle\left\langle k\right\vert $, is realized in an indirect way through coherent state comparison; see the first part of Supplementary Material for details about the PND module. If the projection result is $k=0$, we will obtain the target state
\begin{align}
\left\vert \Phi\right\rangle   =\left\vert H\right\rangle
_{C}\left\vert \phi_{1}\right\rangle _{1}+\left\vert V\right\rangle _{C}\left\vert \phi
_{2}\right\rangle _{2} \label{eq1}%
\end{align}
where the spatial modes $1$, $2$ of the target photon depend
on the polarizations $\left\vert H\right\rangle $, $\left\vert V\right\rangle $ of the control photon, respectively. 
If $k\neq 0$, on the other hand, there will be the output%
\begin{align}
e^{-ik\left(  \pi/2\right)  }\left\vert H\right\rangle _{C}\left\vert \phi_1\right\rangle _{2} +e^{ik\left(  \pi/2\right)  }\left\vert V\right\rangle _{C}\left\vert \phi_2\right\rangle _{1}.
\end{align}
Since the exact photon number $k$ is known, it is possible to removed the unnecessary phase shifts $e^{-ik(\pi/2)}$ and $e^{ik(\pi/2)}$ by a conditional phase shift $-k\pi$ applied on the upper spatial modes, based on the classically feed-forwarded measurement result $k$. Finally, implementing a swapping between the upper and lower spatial modes transforms the above state to the desired one $\left\vert \Phi\right\rangle$.

\begin{figure}[ptb]
\includegraphics[width=10.7cm]{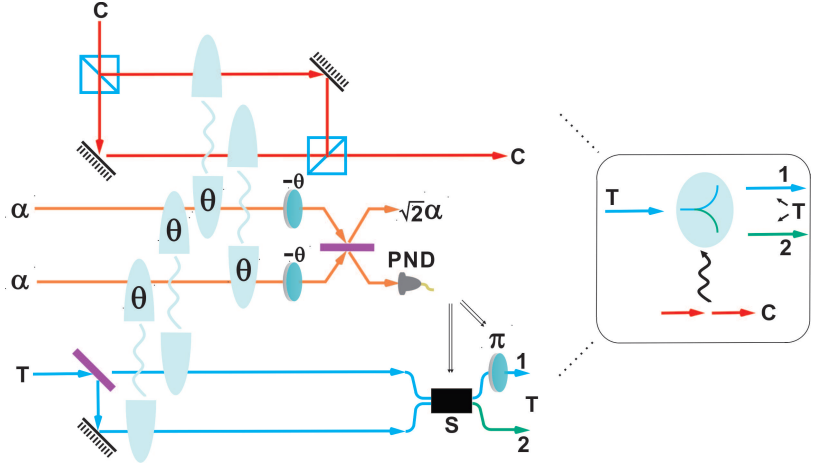}\caption{{Schematic design of an example of c-path gate. The control photon and target photon contain
only one spatial modes. Firstly, the control photon goes through a PBS, and the target photon through a 50:50 BS.
Next, the spatial modes interact with the qubus beam as indicated. The operation steps in the following order---phase shift $-\theta$, 
the two coherent states interference, and the detection of the first coherent-state component by a photon
number-resolving detector (PND) for controlling the switch and phase
shift $\pi$---implements this c-path gate.}}%
\end{figure}

After implementing the c-path gate, the target photon will be separated into two spatial modes, which depend on the polarizations of the control photon. Then a more general c-path gate realizing multiple path mode control should be used for further processing. In a general case, the input state can be given as follows:
\begin{align}
\left\vert \Psi\right\rangle _{n}  &  =\left\vert H\right\rangle _{C^{1}%
,C^{2},\cdots,C^{m}}\left\vert \phi_{1}\right\rangle _{1,2,\cdots
,n}+\left\vert V\right\rangle _{C^{1},C^{2},\cdots,C^{m}}\left\vert \phi
_{2}\right\rangle _{1,2,\cdots,n}, \label{in2} 
\end{align}
The control single photon could have $m$ spatial modes ($C^{1},C^{2},\cdots,C^{m}%
$), and the target photon could have $n$ spatial modes ($1,2,\cdots,n$).
For other applications of such c-path gate, the input single photon states for a general $m$-control-$n$ c-path gate can be directly prepared with linear optical circuits \cite{linear1, linear2}. Through the similar procedure as in the previously discussed 
special c-path gate, 
one will obtain the following state before the detection:
\begin{align}
&  \frac{1}{\sqrt{2}}\left\{  \left\vert H\right\rangle _{C^{1},C^{2}%
,\cdots,C^{m}}\left\vert \phi_{1}\right\rangle _{1,3,\ldots,2n-1}\left\vert
0\right\rangle_{cs} \left\vert \sqrt{2}\alpha\right\rangle_{cs} \right. \nonumber\\
&  +\left\vert H\right\rangle _{C^{1},C^{2},\cdots,C^{m}}\left\vert \phi
_{1}\right\rangle _{2,4,\ldots,2n}\left\vert -\beta\right\rangle_{cs} \left\vert
\sqrt{2}\alpha\cos\theta\right\rangle_{cs} \nonumber\\
&  +\left\vert V\right\rangle _{C^{1},C^{2},\cdots,C^{m}}\left\vert \phi
_{2}\right\rangle _{1,3,\ldots,2n-1}\left\vert \beta\right\rangle_{cs} \left\vert
\sqrt{2}\alpha\cos\theta\right\rangle_{cs} \nonumber\\
&  \left.  +\left\vert V\right\rangle _{C^{1},C^{2},\cdots,C^{m}}\left\vert
\phi_{2}\right\rangle _{2,4,\ldots,2n}\left\vert 0\right\rangle_{cs} \left\vert
\sqrt{2}\alpha\right\rangle_{cs} \right\}  .
\end{align}
After the detection by the PND and the corresponding conditional phase shift if necessary, the following desired state can be achieved,
\begin{align}
\left\vert \Phi\right\rangle _{2n}  &  =\left\vert H\right\rangle
_{C^{1},C^{2},\cdots,C^{m}}\left\vert \phi_{1}\right\rangle _{1,3,\ldots
,2n-1}\nonumber\\
&  +\left\vert V\right\rangle _{C^{1},C^{2},\cdots,C^{m}}\left\vert \phi
_{2}\right\rangle _{2,4,\ldots,2n} \label{eq2}%
\end{align}
where the spatial modes $1$, $3$, $\cdots$, $2n-1$ of the target photon depend
on the polarization ($\left\vert H\right\rangle $) of the control photon;
while the other spatial modes $2$, $4$, $\cdots$, $2n$ depend on the
polarization ($\left\vert V\right\rangle $) of the control photon. 

In the above process, each spatial mode of the target qubit will be separated
into two spatial modes, respectively, depending on the bit information of
control qubit. Regarding the number of operations, all spatial modes of control
and target single photon should interact with the qubus beams, 
necessitating $2n+2m$ XPM processes in total. During an operation, the qubus
beams are not destroyed, i.e. $\left\vert \sqrt{2}\alpha\cos\theta\right\rangle_{cs}
\sim\left\vert \sqrt{2}\alpha\right\rangle_{cs} $, since $\theta$ is tiny. So they
can be reused in the following operations. As discussed in the first part of Supplementary Material, these qubus beams can be used for more than
$10^4$ times, even with a moderate strength, e.g. $|\alpha| \sim 10^3$, and a small
cross phase shift, e.g. $\theta \sim 0.01$. Moreover, one could reduce the number of XPM process to
$n+m$ by moving all interactions into one arm (see
the second part of Supplementary Material for the details). Throughout an operation, only two coherent-state components should be consumed in 
detection. If $n$ and $m$ are very small, we may choose to
save the qubus beams. The alternative is to lower the
amount of XPM processes, given large $n$ or $m$.

\subsection*{Merging gate}
\begin{figure}[ptb]
\includegraphics[width=10.7cm]{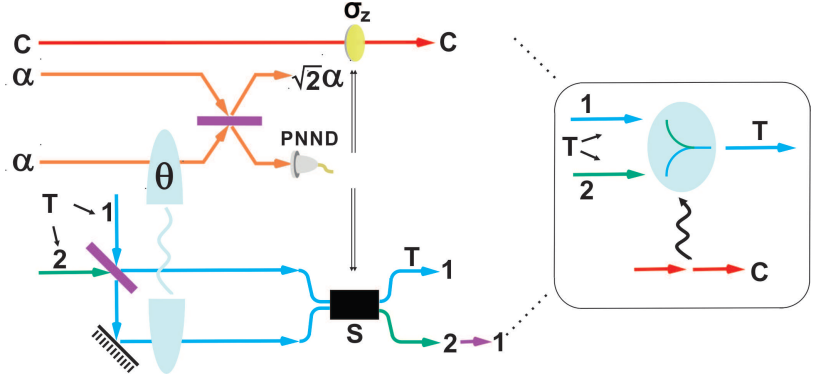}\caption{{Schematic design of an example of merging gate. Firstly, the two spatial modes of the target photon undergo the interference via a 50:50 BS. The spatial mode $2$ will interact with one of the
coherent state. After that, the detection on the first coherent-state component is used to control
the switch and the Pauli operation $\sigma_{z}$, realizing the merging gate.
}}%
\end{figure}

A merging gate performs the inverse transformation of a c-path gate. A special example of the gate for processing the quantum state from the input state in Eq. (\ref{eq1}) is shown in Fig. 2. Firstly, the target photon with
$2$ spatial modes is injected into a 50:50 BS, and then the input state
$\left\vert \Phi\right\rangle$ is transformed to
\begin{align}
&  \frac{1}{\sqrt{2}}\left\{  \left\vert H\right\rangle _{C}\left\vert \phi_{1}\right\rangle _{1}+\left\vert H\right\rangle _{C}\left\vert \phi
_{1}\right\rangle _{2}+\left\vert V\right\rangle _{C}\left\vert \phi_{2}\right\rangle _{1} -\left\vert V\right\rangle _{C}\left\vert
\phi_{2}\right\rangle _{2}\right\}  .
\end{align}
Letting the spatial mode $2$ interact with the qubus beam as shown in Fig.2,
we will get the following state:
\begin{align}
&  \frac{1}{\sqrt{2}}\left\{  \left\vert H\right\rangle _{C}\left\vert \phi_{1}\right\rangle _{1}\left\vert
\alpha\right\rangle_{cs} \left\vert \alpha\right\rangle_{cs} \right. +\left\vert H\right\rangle _{C}\left\vert \phi
_{1}\right\rangle _{2}\left\vert \alpha\right\rangle_{cs} \left\vert
\alpha e^{i\theta}\right\rangle_{cs} \nonumber\\
&  +\left\vert V\right\rangle _{C}\left\vert \phi
_{2}\right\rangle _{1}\left\vert \alpha\right\rangle_{cs} \left\vert
\alpha\right\rangle_{cs}   \left.  -\left\vert V\right\rangle _{C}\left\vert
\phi_{2}\right\rangle _{2}\left\vert \alpha\right\rangle_{cs}
\left\vert \alpha e^{i\theta}\right\rangle_{cs} \right\}  .
\end{align}
After that, one more 50:50 BS and the detection with the result $k=0$ on the first qubus
beam will project the above state into the desired state $\left\vert
\Psi\right\rangle$. Meanwhile, the detection with the result $k\neq0$ will project the
state to $\left\vert H\right\rangle _{C}\left\vert \phi_1
\right\rangle _{2}-\left\vert V\right\rangle _{C}\left\vert \phi_2\right\rangle _{2}$,
where the index $2$ can be redefined as $1$.
Finally, one $\sigma_{z}$ operation on the control photon and a switch of the
upper and lower spatial modes, which are based on the classically feed-forwarded
measurement, will transform the above state to the desired one $\left\vert \Psi\right\rangle$.

Generalized to the case for more than one spatial mode of the control and target photon is straightforward. With the same setups, the initial state as the form of $\left\vert \Phi\right\rangle_{2n}$ can be transformed back to the desired state as the form of $\left\vert \Psi\right\rangle _{n}$. In one word, a merging gate depicted in the above merges the spatial modes of a target photon without
changing anything else. Here, no ancilla single photon is required as
compared with the design in the previous works \cite{qlin2,qlin3,qlin4}. Moreover, only the
discrimination of the vacuum state $\left\vert 0\right\rangle_{cs} $ and the coherent state $\left\vert
\frac{\alpha-\alpha e^{i\theta}}{\sqrt{2}}\right\rangle_{cs} $ is necessary here, with their overlap being in the approximate order 
of $\exp\{-\alpha^2\theta^2/4\}$ to have the close to ideal discrimination of these two states. 
Similar to the operation of the PND module discussed in the first part of Supplementary Material, the discrimination of
the vacuum and coherent state could be realized by a photon number non-resolving detector (PNND) with less than unit efficiency 
$\eta <1$. The error probability of the operation is in the approximate order $\exp\{-\eta\alpha^2\theta^2/2\}$, demanding 
a moderate requirement $\alpha^2\theta^2 \gg 1$. In contrast, the requirements in Homodyne detection scenarios are much tougher; $\alpha\theta^2\gg1$ for the $\hat{X}$-quadrature measurement \cite{Barrett, munro, req1, req2}, and $\alpha\theta\gg1$  for the $\hat{P}$-quadrature measurement \cite{req1, req2}. Similar to the use of qubus beam in the PND module discussed in the first part of Supplementary Material, the
remaining qubus beam is almost the same as the initial one $\left\vert
\frac{\alpha+\alpha e^{i\theta}}{\sqrt{2}}\right\rangle_{cs} \sim\left\vert
\sqrt{2}\alpha\right\rangle_{cs} $, so the qubus beam could be recycled for large number of times as well.
Moreover, only half of the spatial modes of target photon should interact with the qubus beam, demanding only $n$ XPM operations.

\section*{Multi-control unitary operations and general unitary operation}

Since the combination of a pair of c-path gate and merging gate (associated
with a bit flip operation) can be used to realize a CNOT gate, these two
element gates are universal for circuit-based quantum computation
\cite{qlin2}. In addition, we will show that c-path and merging
gate make it possible to realize various controlled unitary operations involving large number of qubits in 
more efficient way.

\subsection*{general (\textit{n }-- 1)-control-1 unitary operation}

The first gate is a general $(n-1)$-control-$1$ gate called uniform
controlled rotation \cite{Mottonen} or multiplexor \cite{Shende2}. 
It implements an operation represented by the following matrix:%
\begin{equation}
F_{1,n}^{n-1}=\left(
\begin{array}
[c]{cccc}%
U^{(1)}_{1} &  &  & \\
& U^{(1)}_{2} &  & \\
&  & \vdots & \\
&  &  & U^{(1)}_{2^{n-1}}%
\end{array}
\right)  ,
\end{equation}
where the subscripts $\left(  1,n\right)$ means that the target qubit
is the $n$-th qubit, and the superscript $\left(  n-1\right)$ indicates that the
control qubits are the other $n-1$ qubits. $U^{(1)}_{i}\left(  i=1,\cdots
,2^{n-1}\right)$ in the matrix are the single-qubit unitary operations. With $U^{(1)}_{i}=\mathbb{I},$
for $i=1,\cdots,2^{n-1}-1$, and $U^{(1)}_{2^{n-1}}=\sigma_{x}$, the gate is a $\left(
n-1\right)  $-controlled Toffoli gate. To realize a Toffoli gate,
$\mathcal{O}\left(  n^{2}\right)  $ CNOT gates should be necessary
\cite{Barenco}, while for a general $(n-1)$-control-$1$ gate, the required CNOT
gate number should be increased to $2^{n-1}$ \cite{Mottonen}. However, 
the complexity to realize a Toffoli gate can be reduced to linear by using $n-1$ pairs of c-path and
merging gates \cite{qlin2, qlin3}. Here we will show that this approach can be generalized to realize a 
general $(n-1)$-control-$1$ gate efficiently.

For clarity, we use the example of $3$-control-$1$ unitary operation
for illustration (see Fig.3). The input state can be the following general
4-qubit state, $\underset{i=1}{\overset{8}{\sum}}\left\vert i\right\rangle ^{123}%
\left\vert \phi_{i}\right\rangle ^{4}$,
where $\left\vert 1\right\rangle ^{123}=\left\vert HHH\right\rangle $,
$\left\vert 2\right\rangle ^{123}=\left\vert HHV\right\rangle ,$ etc. and
$\left\vert \phi_{i}\right\rangle =\alpha_{i}\left\vert H\right\rangle
+\beta_{i}\left\vert V\right\rangle $ with $\sum\limits_{i=1}^{8}\left(
\left\vert \alpha_{i}\right\vert ^{2}+\left\vert \beta_{i}\right\vert
^{2}\right)  =1$. As shown in Fig. 3, with the sequential operation of three c-path gates on the
target photon, the input state will be transformed to $\underset{i=1}{\overset{8}{\sum}}\left\vert i\right\rangle ^{123}%
\left\vert \phi_{i}\right\rangle _{i}^{4}$, where the subscript $i$ outside the bracket denotes the spatial modes of the
fourth qubit. The eight spatial modes are determined by the bit information
of the other three control qubits. If eight single-qubit unitary operations
($U^{(1)}_{i},i=1,\cdots,8$) are preformed on the corresponding spatial modes, one
will obtain the state, $\underset{i=1}{\overset{8}{\sum}}\left\vert i\right\rangle ^{123}%
U^{(1)}_{i}\left\vert \phi_{i}\right\rangle _{i}^{4}$.
Finally, after three merging gates erase the path information of the spatial modes $i$, the
desired state $\underset{i=1}{\overset{8}{\sum}}\left\vert i\right\rangle
^{123}U^{(1)}_{i}\left\vert \phi_{i}\right\rangle ^{4}$ will be achieved, realizing
the $3$-control-$1$ unitary operation. Generalizing to $n$-qubit case is
straightforward with $(n-1)$ pairs of c-path and merging gates. It is
significantly simpler than the traditional CNOT-based approach, which demands 
complicated decomposition into exponentially large number CNOT gates.

\begin{figure}[ptb]
\includegraphics[width=12.7cm]{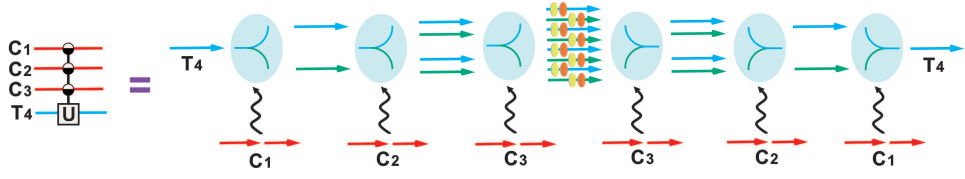}\caption{{Realization of
$3$-control-$1$ unitary operation. By three c-path gates controlled by the
photons $C_{1}, C_{2}, C_{3}$ sequentially, the target photon will be
separated into $8$ spatial modes. After the single-photon unitary
operations on the corresponding spatial modes, this unitary operation will be realized, 
associated with the operations of three merging gates.
}}%
\end{figure}

The quantity of XPM process for implementing the linearly scaling c-path and merging gate pairs is 
also an indicator for the complexity of the gate designs. By applying the c-path gates step by step, the involved
spatial modes of the target photon will be exponentially increased 
with the operation procedure. Generally, for the $m$-th c-path gate, the target photon will be separated
into $2^{m}$ spatial modes, which should be coupled to the qubus beam.
Therefore, $2^{m-1}+1$ XPM processes are required in the $m$-th c-path gate (here
the c-path gate is the modified one discussed in the second part of Supplementary Material). On the other
hand, for the inverse $m$-th merging gate (the operation order is from $n-1$ to $1$),
$2^{m-1}$ XPM processes are necessary too. Totally, the number of the necessary XPM process
for the realization of a $(n-1)$-control-$1$ gate should be%

\begin{equation}
N_n=\underset{m=1}{\overset{n-1}{\sum}}\left(  2^{m-1}+1+2^{m-1}\right)
=2^{n}+n-3.
\end{equation}
If the modified c-path gates are replaced by the original c-path gates
discussed before, the required number of XPM process will be increased to
\begin{equation}
N_n=\underset{m=1}{\overset{n-1}{\sum}}\left(  2^{m}+2+2^{m-1}\right)
=3\cdot2^{n-1}+2n-5. 
\end{equation}
This exponential increasing is due to the fact that the general
$(n-1)$-control-$1$ gate itself is exponential complexity. Since there 
are exponential $2^{n-1}$ control unitary operations in such gate, the exponentially increasing 
number of XPM process will be inevitable.

In addition to the number of XPM process, the amount of other
operations such as single photon interference, coherent state interference,
as well as the required resources such as qubus beams (not including
ancilla single photons), are only linearly increasing with the involved photonic 
qubit number. That is a considerable improvement over the former CNOT-based
approach, which requires the amount of interferences, qubus beams, ancilla
single photons and others in the exponential orders. In this sense, our current approach provides a more feasible way to
realize a general $(n-1)$-control-$1$ unitary operation.

\subsection*{special (\textit{n }-- 1)-control-1 unitary operation}

It is possible to optimize the implementation of the following special
$(n-1)$-control-1 gate operation

\begin{equation}
F_{s,1,n}^{n-1}=\left(
\begin{array}
[c]{cccccc}%
\mathbb{I} &  &  &  &  & \\
& \vdots &  &  &  & \\
&  & \mathbb{I} &  &  & \\
&  &  & U^{(1)}_{2^{n-1}-2^{n-m-1}+1} &  & \\
&  &  &  & \vdots & \\
&  &  &  &  & U^{(1)}_{2^{n-1}}%
\end{array}
\right)  , \label{sc}%
\end{equation}
where $m\leq n-1.$ In this operation the target
photon will not be affected when one of the first $m$ control photons is in
the state $\left\vert H\right\rangle $, but it will be
under operation when the first control photons are all in the state $\left\vert
V\right\rangle $.

Without loss of generality we use a 5-qubit gate in Fig. 4 as example. Here we briefly describe steps, 
and the details can be found in the third part of Supplementary Material. At first, the photon $C_{1}$ will
control the photon $C_{2}$, but does not act on the target photon directly. Next, the
three spatial modes ($1,1^{\prime},2$) of the photon $2$ will control all rest photons, including
the target photon. After that, the photon $C_{3}, C_{4}, T_{5}$ will be
separated into two spatial modes $\left(  1,2\right) $, respectively. Applying
the general $2$-control-$1$ unitary operation on the spatial mode $2$ of the
photon $C_{3}, C_{4}, T_{5}$, the desired 5-qubit gate will be completed
associated with the corresponding merging gates.

\begin{figure}[ptb]
\includegraphics[width=12.7cm]{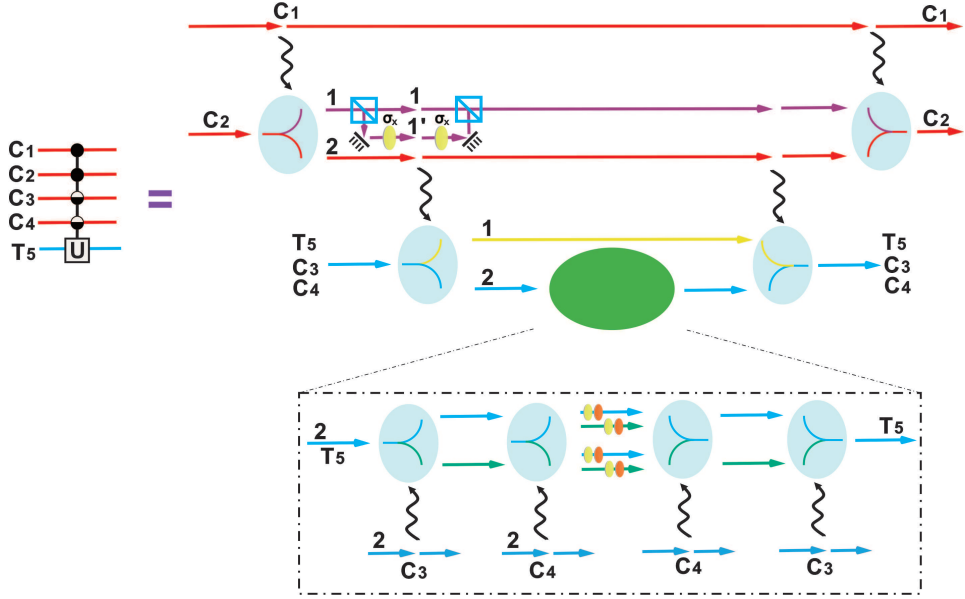}\caption{{Procedure of implementing 
the $4$-control-$1$ unitary operation outlined in the left panel. 
Step one: photon $1$ controls photon $2$ by the first c-path gate. Step two: the
spatial mode $1$ passes through a PBS and a $\sigma_{x}$ operation to
the $\left\vert V\right\rangle $ mode is applied. Step three: the three spatial modes are
used as the control modes to control the other three photons $3,4,5$ by three
c-path gates. Final step: a general $2$-control-$1$
operation to the spatial modes 2 of photons $3,4,5$ is applied, together with the inverse merging gate operations.
}}%
\end{figure}

Generalizing to the unitary operation of Eq. (\ref{sc}) is
straightforward. Firstly, $m-1$ c-path operations are performed to the first
$m$ photons in turn. After that, using the spatial modes ($1,1^{\prime},2$) of
the $m$-th control photon as the control modes for the following $n-m$ c-path
gates, all of the rest photons including the target photon will be separated into two
spatial modes $\left(  1,2\right)  $, respectively. Finally, by applying the
general ($n-m-1$)-control-$1$ unitary operation discussed before to the
spatial modes $2$ of the rest photons, associated with the corresponding $n-1$
merging gates, the desired unitary operation will be completed.

Now we discuss the complexity of the procedure. The first modified c-path gate requires $2$ XPM
processes, and each of the other $n-2$ modified c-path gates needs $2$ XPM
processes as well (the control photon has three spatial modes and two of them
are in the state $\left\vert H\right\rangle $, which will not interact with
the qubus beam). For a general ($n-m-1$)-control-$1$ gate, $2^{n-m}+n-m-3$
XPM processes are required. Meanwhile, for the $n-1$ merging gate, $n-1$ XPM
processes are sufficient. The total number of XPM processes should be $2^{n-m}+4n-m-6$.
Compared with the amount of the general $n$-control-$1$ gate, it is a
considerable improvement by reducing a factor of $2^{m}$. Obviously, it is on the same order of amount as 
the non-identity operations in Eq. (\ref{sc}). Especially, if
$m=n-1$, it will be a general Toffoli gate, which can be implemented with $3n-3$ XPM processes scaling linearly with $n$.

\subsection*{1-control-(\textit{n }-- 1) unitary operation}

\begin{figure}[ptb]
\includegraphics[width=9.7cm]{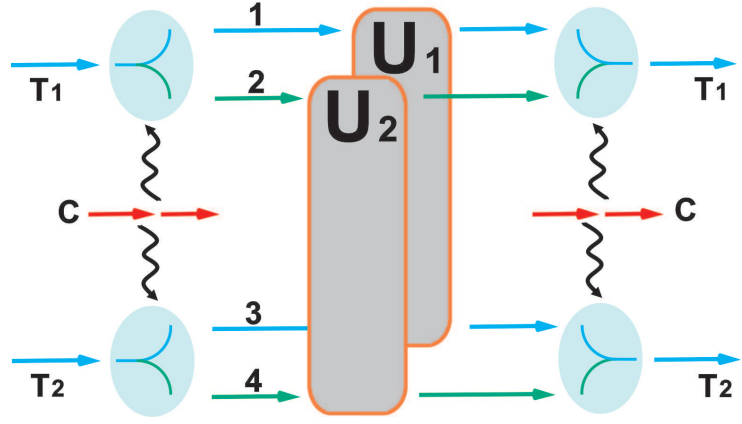}\caption{{Implementation of
$1$-control-$2$ unitary operation. Firstly, apply two c-path gate operations between the
control photon and the two target photons, respectively. After that, two
unitary operations are applied to the corresponding spatial modes as shown in
the figure. Finally, the whole procedure will be completed by two merging
gates.
}}%
\end{figure}

Now, we consider another $n$-qubit gate, through which one qubit controls the other
$n-1$ qubits. Its operation is described by the matrix
\begin{equation}
F_{n-1}^{1}=\left(
\begin{array}
[c]{cc}%
U_{1}^{(n-1)} & \\
& U_{2}^{(n-1)}%
\end{array}
\right)  ,
\end{equation}
where $U_{1(2)}^{(n-1)}$ denotes a $\left(  n-1\right)  $-qubit unitary
operation. We first consider the example of $1$-control-$2$ gate (see Fig.5). 
This $1$-control-$2$ unitary operation implements unitary operations $U_{1}^{(2)},U_{2}^{(2)}$ on
two target qubits when the control qubit in the states $\left\vert
H\right\rangle ,\left\vert V\right\rangle $, respectively. Suppose that the input
state is $\left\vert H\right\rangle _{C}\otimes\left\vert \psi\right\rangle
_{T_{1}T_{2}}+\left\vert V\right\rangle _{C}\otimes\left\vert \varphi
\right\rangle _{T_{1}T_{2}}$, where $\left\vert \psi\right\rangle _{T_{1}%
T_{2}}$ and $\left\vert \varphi\right\rangle _{T_{1}T_{2}}$ can be in
arbitrary forms. Firstly, one uses two c-path gates to separate the two 
target qubits into two spatial modes ($1,2$) or ($3,4$), respectively, i.e. 
the obtained state is $\left\vert H\right\rangle _{C}\otimes\left\vert \psi\right\rangle
_{13}+\left\vert V\right\rangle _{C}\otimes\left\vert \varphi\right\rangle
_{24}$. After that, implementing the desired unitary operations on the two spatial
modes ($2,3$) or ($1,4$), respectively, will yield the following state%
\begin{equation}
\left\vert H\right\rangle _{C}\otimes U_{1}^{(2)}\left\vert \psi\right\rangle
_{13}+\left\vert V\right\rangle _{C}\otimes U_{2}^{(2)}\left\vert
\varphi\right\rangle _{24}.
\end{equation}
Finally, two merging gates are used to erase the path information, and then 
one will achieve the target state
\begin{equation}
\left\vert H\right\rangle _{C}\otimes U_{1}^{(2)}\left\vert \psi\right\rangle
_{T_{1}T_{2}}+\left\vert V\right\rangle _{C}\otimes U_{2}^{(2)}\left\vert
\varphi\right\rangle _{T_{1}T_{2}}.
\end{equation}
Here, two pairs of c-path and merging gate, associated with two two-qubit
unitary operations, will be needed. The required sources are also obviously
fewer than the CNOT-based approach. Especially, if $U_{1}^{(2)}=\mathbb{I}$
 and $U_{2}^{(2)}=$ SWAP, it will be a Fredkin gate. In the CNOT-based approach this gate 
requires 5 CNOT gates \cite{Smolin} (one CNOT is equivalent to a pair of c-path and merging gate, or
two parity-check operations \cite{Pittman}), while only two pairs of c-path
and merging gates (associated with a spatial mode swap operation) are necessary to construct the 
gate \cite{qlin2, qlin3}.

Its generalization is straightforward with three similar
processes will complete the operation; (1) $n-1$ c-path gates
separate each of the $n-1$ target qubits into two spatial modes; (2) the unitary operation $U_{1}^{(n-1)}$ 
is performed on the spatial modes corresponding to the state $\left\vert H\right\rangle $ of the control qubit, 
and the unitary operation $U_{2}^{(n-1)}$ on the other spatial modes
simultaneously; (3) $n-1$ merging gates will merge the target qubits. 
Totally, except for the requirement for
realizing unitary operations $U_{1}^{(n-1)}$ and $U_{2}^{(n-1)}$, the required
sources increase linearly with involved qubits number ($n-1$ pairs of c-path
and merging gates). \

\subsection*{\textit{n}-control-\textit{m} unitary operation}

In what follows, we will discuss the $n$-control-$m$ unitary operation, which is
described by the following:%

\begin{equation}
F_{m}^{n}=\left(
\begin{array}
[c]{cccc}%
U_{1}^{(m)} &  &  & \\
& U_{2}^{(m)} &  & \\
&  & \vdots & \\
&  &  & U_{2^{n-1}}^{(m)}%
\end{array}
\right)  ,
\end{equation}
where $U_{i}^{(m)}$ $(i=1,\cdots,2^{n-1})$ are the $m$-qubit unitary
operations. We combine the structures of the above gates to realize
this $n$-control-$m$ unitary operation. Firstly, we use $n$ c-path gates to
separate the first target qubit into $2^{n-1}$ spatial modes, and then use 
$n$ c-path gates to separate the other target qubits,
respectively. After that, we implement the unitary operation $U_{1}^{(m)}$ on
the first spatial modes of each target qubits, 
followed by implementing the other $m$-qubit unitary operations $U_{i}^{(m)}$ to the corresponding
spatial modes. Finally, the merging gates will merge the target qubits
to complete the whole procedure. Totally $n\times m$ pairs of c-path and
merging gate are required, in addition to the $m$-qubit unitary operations.

\subsection*{General $n$-qubit unitary operation}

The most general operation is $n$-qubit unitary operation. This operation
is a crucial operation in quantum information processing, since it simulate the evolution 
of $n$ spin-1/2 interacting particles. This simulation is impossible by
classical computer. A general n-qubit unitary operation has $4^{n}-1$ degrees
of freedom from a $2^{n}\times 2^{n}$ unitary matrix. Numerous
works have been devoted to the problem of how to construct a general $n$-qubit
unitary operation with two-qubit gates and single-qubit gates \cite{Knill,
Cybenko, Aho, Shende1, Vartiainen, Shende2}. The theoretical lower bound of
the CNOT approach is $\frac{1}{4}\left(  4^{n}-3n-1\right)  $ \cite{Shende1}.
However, it is only a theoretical limit, and the detailed construction of 
such circuit had not discovered yet. The best circuit construction is the quantum Shannon
decomposition (QSD), using $\left(  23/48\right)  \times4^{n}-\left(
3/2\right)  \times2^{n}+4/3$ CNOT gates \cite{Shende2}. Here we will present
two approaches to the construction of a general $n$-qubit unitary operation with c-path
and merging gates.

\subsubsection*{approach based on cosine-sine decomposition}

The first approach is based on the cosine-sine decomposition (CSD)
\cite{csc1,csc2}. By this method a general $n$-qubit unitary operation can be
decomposed into the following form:%

\begin{equation}
U^{\left(  n\right)  }=F_{n-1}^{1}\left(
\begin{array}
[c]{cc}%
\mathbb{C}^{\left(  n-1\right)  } & -\mathbb{S}^{\left(  n-1\right)  }\\
\mathbb{S}^{\left(  n-1\right)  } & \mathbb{C}^{\left(  n-1\right)  }%
\end{array}
\right)  F_{n-1}^{\prime1},
\end{equation}
where $F_{n-1}^{1},F_{n-1}^{\prime1}$ are the 1-control-$(n-1)$ unitary
operations, and $\mathbb{C}^{\left(  n-1\right)  },\mathbb{S}^{\left(  n-1\right)  }$ are the
real diagonal matrices satisfying $\left[  \mathbb{C}^{\left(  n-1\right)  }\right]
^{2}+\left[  \mathbb{S}^{\left(  n-1\right)  }\right]  ^{2}=\mathbb{I.}$  It has been
demonstrated that the middle operation is equivalent to a $(n-1)$-control-1
unitary operation \cite{Shende2}. With the above decomposition it is evident
that one could combine two 1-control-$(n-1)$ unitary
operations and one $(n-1)$-control-1 unitary operation to realize a general
$n$-qubit unitary operation. Therefore we will get the following recursive relation%
\begin{equation}
N_{n}=4N_{n-1}+\left(  2^{n}+n-3\right)  +6\left(  n-1\right)  ,
\end{equation}
for the number of XPM process, where the second term is the amount
of the middle $(n-1)$-control-1 unitary operation, and the third is that of 
the c-path and merging gates used in the first turn of 1-control-$(n-1)$ gate. Then, 
the total number of XPM process is found as
\begin{equation}
N_{n}=10/9\times4^{n}-2^{n}-7/3\times n-1/9. \label{nu}%
\end{equation}

\subsubsection*{approach based on further decomposition into general
$(n-1)$-control-$1$ unitary operations}

One can also decompose a general unitary operation without CSD. It was demonstrated that a general n-qubit unitary
operation can be decomposed into a series of $(n-1)$-control-$1$ unitary
operations as follows \cite{Mottonen}:%

\begin{equation}
U^{\left(  n\right)  }=\overset{2^{n-1}-1}{\prod\limits_{j=1}}\left(
F_{1,n}^{n-1}\times F_{1,\gamma\left(  j\right)  -1}^{n-1}\right)
_{j}\overset{n-1}{\prod\limits_{i=1}}F_{1,n-i+1}^{n-i},
\end{equation}
where the function $\gamma\left(  j\right)  $ indicates the position of the
least significant nonzero bit in the $n$-bit binary presentation of the number
$j$. Obviously, the above decomposition allows the realization of general
unitary operation, together with a general $(n-1)$-control-$1$ unitary
operation discussed before. There are $2\left(  2^{n-1}-1\right)  $ general $(n-1)$-control-$1$ 
unitary operations ($F_{1,n}^{n-1},F_{1,\gamma\left(  j\right)  -1}^{n-1}$) and
$\left(  n-1\right)  $ general $(n-i)$-control-$1$ unitary operations
($F_{1,n-i+1}^{n-i}$). Therefore, the total number of the required XPM process
will be%

\begin{align}
N_n & =  2\left(  2^{n-1}-1\right)  \times\left(  2^{n}+n-3\right) \label{sx} +\overset{n-1}{\sum\limits_{i=1}}\left(  2^{n-i+1}+n-i+1-3\right)
\nonumber\\
&  =4^{n}+(n-3)\times2^{n}+\frac{1}{2}\left(  n^{2}-9n+8\right)  .\nonumber
\end{align}

\subsubsection*{comparison between complexity}

Now we compare our approaches with the CNOT-based approach in terms of their complexity. 
An optical CNOT gate demands two parity-check and one single photon as ancilla
\cite{Pittman}. If assisted with weak cross-Kerr nonlinearity, two XPM
processes will be needed for one parity-check \cite{Barrett, munro}. This number
can be reduced to one by saving one qubus beam at the price of lowering the success probability 
by half \cite{wang}. In other words, a CNOT gate requires two XPM processes, associated
with one ancilla single photon in addition to the qubus beams. 
Alternatively one could use more XPM processes and qubus beams to have deterministic operation. 
In this case, a CNOT gate requires four XPM processes involving an ancilla single photon. 
Moreover, the number of
interference processes should be taken into account. A parity-check operation
works with one two-photon interference process and one coherent-state
interference process, implying that a CNOT gate needs four interference processes.

In Table I we list the source requirements of the CNOT approach and our
c-path-merging approach for comparison. There, each rows include two quantities, one for
those using less XPM processes and more qubus beams, and the other for those
using more XPM processes to save qubus beams that could be recycled. The
first row is the theoretical lower bound of CNOT approach. Totally,
$\frac{1}{4}\left(  4^{n}-3n-1\right)  $ CNOT gates are required for a general
unitary operation \cite{Shende1}. The quantities of XPM processes, qubus
beams, ancilla single photons and interference processes are based on the number. 
The second row is about the CNOT-based circuit, with the optimal number $\left(  23/48\right)  \times4^{n}-\left(  3/2\right)  \times2^{n}+4/3$.

The required resources for our first approach based on CSD
are given in the third row. The amount of XPM process for the modified c-path gate is shown in Eq. (\ref{nu}). 
Since the qubus beam will be detected with the
probability $1/2$, the corresponding amount of qubus beams can be calculated
by the recursive relation $A_{n}=4A_{n-1}+3\left(  n-1\right)  /2$. Exactly this is just half of the 
number of c-path-merging pairs used for the gate. Moreover, two interference
processes are necessary in a c-path gate and a merging gate, respectively. 
Totally, $\frac{4}{3}\cdot4^{n}-4n-\frac{4}{3}$
interference processes should be used in our first approach. If using the original
c-path gate shown in Fig.1, the amount of XPM processes will be increased to
the scaling $\mathcal{O}\left(  \frac{11}{6}\times4^{n}\right)  $, which is
found by the relation 
\begin{align}
N_{n}=4N_{n-1}+\left(  3\times2^{n-1}+2n-5\right)
+10\left(  n-1\right). 
\end{align} 
This is smaller than the corresponding number of the CNOT approach. 
Especially, in our approach, no ancilla single photon is necessary in contrast to the CNOT 
approach.

In addition to the amount of XPM processes shown in Tab. I, the corresponding number
of qubus beams is 
\begin{align}
N_{q}&=\left(  2^{n-1}-1\right)  \times\left(  n-1\right)
+\overset{n-1}{\sum\limits_{i=1}}\left(  n-i\right)  /2\nonumber\\
&=\left(  n-1\right)
\cdot2^{n-1}+\left(  n^{2}-5n+4\right)  /4.
\end{align} 
Obviously, this number scales as
$n2^{n-1}$, much lower than those of the three other approaches by a factor 
of $2^{n}/n$. Since this quantity happens to be half of the number of c-path-merging
pairs, the number of required interference processes is eight times of this
number. If using the original c-path gates, the required XPM processes will be
increased to 
\begin{align}
N_n&=2\left(  2^{n-1}-1\right)  \times\left(  3\cdot2^{n-1}%
+2n-5\right)  +\overset{n-1}{\sum\limits_{i=1}}\left(  3\cdot2^{n-i}+2\left(
n-i+1\right)  -5\right).
\end{align} 

Through the comparison it is evident that our approaches enjoy the 
advantages of no ancilla single photons, less ancilla
resources (qubus beams) and fewer operations (interference processes). Our
second approach with the modified c-path gate is the optimal one to 
realize a general unitary operation. It should be noted that the sources for doing the measurements are not taken into account in the above discussion, since we only focus on the complexity of the schemes themselves and the measurements only use more sources of constant amount if they are performed by a few modules (like that described in the first part of Supplementary Material) in succession.

\begin{widetext}
\begin{table}[h]%
\begin{tabular}
[c]{|c|c|c|c|c|}\hline
& XPM processes & qubus beams & ancilla single photons & interference processes\\\hline
& $\frac{1}{2}\left(  4^{n}-3n-1\right)  $ & $\frac{1}{4}\left(
4^{n}-3n-1\right)  $ & $\frac{1}{4}\left(  4^{n}-3n-1\right)  $ & \\
\raisebox{2ex}{CNOT-1\cite{Shende1}} & $4^{n}-3n-1$ & $0$ & $\frac{1}%
{4}\left(  4^{n}-3n-1\right)  $ & \raisebox{2ex}{$4^{n}-3n-1$}\\\hline
& $\frac{23}{24}\cdot4^{n}-3\cdot2^{n}+\frac{8}{3}$ &
$\frac{23}{48}\cdot4^{n}-\frac{3}{2}\cdot2^{n}+\frac{4}{3}$ & $\frac{23}{48}\cdot4^{n}-\frac
{3}{2}2^{n}+\frac{4}{3}$ & \\
\raisebox{2ex}{CNOT-2\cite{Shende2}} & $\frac{23}{12}\cdot4^{n}-6\cdot2^{n}+\frac{16}{3}$ &
$0$ & $\frac{23}{48}\cdot4^{n}-\frac
{3}{2}2^{n}+\frac{4}{3}$ & \raisebox{2ex}{$\frac{23}{12}\cdot4^{n}-6\cdot2^{n}+\frac{16}{3}$}\\\hline
& $\frac{10}{9}\cdot4^{n}-2^{n}-\frac{7n}{3}-\frac{1}{9}$ &
$\frac{1}{6}\cdot4^{n}-\frac{1}{2}\cdot n-\frac{1}{6}$ & $0$ & \\
\raisebox{2ex}{c-path-merging-1} & $\frac{11}{6}\cdot4^{n}-3\cdot2^{n-1}-4n-\frac{1}{3}$ &
$0$ & $0$ & \raisebox{2ex}{$\frac{4}{3}\cdot4^{n}-4n-\frac{4}{3}$}\\\hline
& $4^{n}+(n-3)\cdot2^{n}+\frac{1}{2}\left(  n^{2}-9n+8\right)$ & $(n-1)\cdot2^{n-1}+\frac{n^{2}-5n+4}{4}$ & $0$ & \\
\raisebox{2ex}{c-path-merging-2} & $\frac{3}{2}\cdot4^{n}+(2n-5)\cdot2^{n}+n^{2}-8n+7$ & $0$ & $0$ &\raisebox{2ex}{$4(n-1)\cdot2^{n}+2(n^{2}-5n+4)$} \\\hline
\end{tabular}
\caption{Comparison of CNOT approach and c-path-merging approach. The first and second rows are the theoretical lower bound 
and
the known-circuit of CNOT-based approach, respectively, while the third and fourth rows are for the two c-path-merging approaches based on
CSD directly and further decomposition, respectively.}%
\end{table}
\end{widetext}

\section*{Discussion on feasibility of XPM}

The crucial element in our approach is the XPM based on Kerr nonlinearity. Here we approximate the XPM as a single-mode process. In reality, however, photons carry continuous frequency distributions, and the multi-mode character can affect an XPM process. 
In view of the phase noise existing in non-instantaneous Kerr nonlinearity \cite{pnoise}, a multi-mode effect induced imperfection of XPM was first considered by Shapiro and collaborators \cite{Shap, Shapiro}. They conclude that the phase noise due to the non-instantaneous response of Kerr medium can impair the ideal operation of XPM. The non-instantaneous response to light field can happen in optical fiber of silicon and other similar materials. With their extremely small Kerr coefficients, a considerably lengthy fiber should be used to generate a sufficient nonlinear phase. However, a dominant process in fiber is the absorption of the light, which leads to the decoherence of the generated photonic states \cite{real}. This essential point excludes the feasibility of the setups that are relevant to the phase noise problem. On the other hand, the systems that realize much higher 
Kerr coefficients are the coherently prepared atomic ensembles under the conditions of electromagnetically induced transparency (EIT). The response of these atomic ensembles to the input light field is virtually immediate in activating the third and higher order nonlinearity as demonstrated by the experimental \cite{ex1, ex2, ex3} and theoretical studies \cite{julio, rydberg, BECx,  xpm-01, xpm-02, photon}, thus neglecting the phase noise effect in such Kerr nonlinearities. 

Another imperfections due to the multi-mode nature of inputs is the mode entanglement under photonic coupling or interaction 
\cite{julio, rydberg, xpm-01, xpm-02, photon}. Relevant to the Kerr nonlinearity based on atomic ensembles, this effect deviates a real XPM process from the ideal one $|1\rangle|1\rangle\rightarrow e^{i\theta}|1\rangle|1\rangle$ for a pair of single photon states and $|1\rangle|\alpha\rangle_{cs}\rightarrow |1\rangle|e^{i\theta}\alpha\rangle_{cs}$ between a single photon and a coherent state.
For the XPM considered in this paper, it is possible to eliminate the mode entanglement by adopting the counter-propagation configuration and transverse confinement of the inputs \cite{xpm-01}, so that a close to single-mode XPM will be possible 
in a normal EIT medium. The improvement on the intensity of Kerr nonlinearity is feasible via the non-local atomic interaction in other atomic ensembles \cite{photon}. Currently both experimental and theoretical progress toward practical Kerr nonlinearity are under way.  

\section*{Conclusion}

With the improved designs of c-path and merging gate, the realization
of various control unitary operations can be more efficient and with less sources and operations. Compared with the widely considered approach based on CNOT gate and the previously proposed schemes based 
on the original c-path and merging gate, the improvement on the designs of some multi-qubit gates is significant.
For example, a general $n$-control-$1$ gate can be realized by linearly increasing pairs of c-path and merging gate with the number 
of processed photons, while no ancilla photon is needed in operations.
The close to ideal XPM process used in the circuits, as well as in the detection module, would be available with the development of the techniques of Kerr nonlinearity. Based on this prerequisite, the schemes proposed in the current study could become competitive alternatives for large scale photonic quantum computation with their considerably relaxed requirements on sources and operation times.

\begin{acknowledgments}
The authors thank Ru-Bing Yang for helpful suggestions. Q. L. was funded by
National Natural Science Foundation of China (Grant No.11005040), Natural Science Foundation of Fujian Province of China (Grant No. 2014J01015), Program for
New Century Excellent Talents in Fujian Province University (Grant No.
2012FJ-NCET-ZR04) and Promotion Program for Young and Middle-aged Teacher in
Science and Technology Research of Huaqiao University (Grant No. ZQN-PY113).
\end{acknowledgments}

\bigskip

\bf{Contributions}

Q. L. and B. H. designed the scheme. Q. L. prepared the display items (figures and tables). Q. L. and B. H. co-wrote the manuscript.

\bigskip

\bf{Competing financial interests}

The authors declare no competing financial interests.

\end{document}


\title{Supplementary Material for "Highly Efficient Processing Multi-photon States''}
\author{Qing Lin}
\email{qlin@mail.ustc.edu.cn}
\affiliation{Fujian Provincial Key Laboratory of Light Propagation and Transformation, College of Information Science and Engineering, Huaqiao University,
Xiamen 361021, China}
\author{Bing He}
\email{binghe@uark.edu}
\affiliation{Department of Physics, University of Arkansas, Fayetteville, AR 72701, USA}
\maketitle

\vspace{0cm}  

\vspace{0cm}  

\subsection*{The photon number-resolving detection module}
\renewcommand{\theequation}{A-\arabic{equation}}
\renewcommand{\thefigure}{A-\arabic{figure}} 
\setcounter{equation}{0}
\setcounter{figure}{0} 

We discuss the realization of the photon number-resolving detector (PND) in this part. The idea of our design is to realize the PND in an indirect way with photon number non-resolving detector (of less than unit efficiency) and coherent state comparison. More details of the proposal can be found in \cite{qlin2, qlin3}. Here we present an outline of the implementation of the PND module with Fig. A-1. Suppose that the photon number non-resolving detector used in the module has the efficiency $\eta<1$. The aim is to resolve the exact Fock state components in the coherent state component $\left \vert \pm\beta\right\rangle_{cs}=\left\vert \pm i\sqrt{2}\alpha\sin\theta\right\rangle_{cs}$ with $\theta\ll 1$. We let it interacted with one of the two qubus beams $\left\vert\gamma\right\rangle_{cs}\left\vert\gamma\right\rangle_{cs}$ through an XPM process. The transformation of the qubus beam state is as follows,
\begin{align}
\left \vert \pm\beta\right\rangle_{cs}\left\vert\gamma\right\rangle_{cs}\left\vert\gamma\right\rangle_{cs} \rightarrow e^{-\left|\beta\right|^2/2} \sum_{k=0}^{\infty }\frac{\left( \pm \beta\right)^k}{\sqrt{k!}}\left\vert k\right\rangle \left\vert \gamma e^{ik\theta}\right\rangle_{cs}\left\vert\gamma\right\rangle_{cs}.
\end{align}
After that, one more beam splitter (BS) is applied to get the state
\begin{align}
e^{-\left|\beta\right|^2/2} \sum_{k=0}^{\infty }\frac{\left( \pm \beta\right)^k}{\sqrt{k!}}\left\vert k\right\rangle \left\vert \frac{\gamma e^{ik\theta}-\gamma}{\sqrt{2}}\right\rangle_{cs}\left\vert \frac{\gamma e^{ik\theta}+\gamma}{\sqrt{2}}\right\rangle_{cs}.
\end{align}
Now the information of the Fock state components in $\left\vert \pm \beta\right\rangle_{cs}$ has been contained in the coherent state $\left\vert \frac{\gamma e^{ik\theta}-\gamma}{\sqrt{2}}\right\rangle_{cs}$. 

For the different numbers $k$, the Poisson distributions of the coherent states $\left\vert \frac{\gamma e^{ik\theta}-\gamma}{\sqrt{2}}\right\rangle_{cs}$, for $k=1,2,\dots$, can be well separated with the approximate overlap 
$\exp{\{-\gamma^2 \theta^2/4\}}$, given a sufficiently large $|\gamma|$; see Fig. A-1. It makes the discrimination of 
the coherent states $\left\vert \frac{\gamma e^{ik\theta}-\gamma}{\sqrt{2}}\right\rangle_{cs}$ for the different $k$ possible. 
The response of a photon number non-solving detector to these coherent states, which occur with the corresponding probabilities 
as in the above equation, will be distinct. Moreover, due to the the small $|\beta|$, the number of the Poisson peaks in Fig. A-1 is limited. Thus, the projection $|k\rangle\langle k|$ on the coherent $ \left \vert \pm\beta\right\rangle_{cs}$ will be 
realized indirectly with the module in Fig. A-1. 

The operation of the photon number non-resolving detector can be described by the following positive operator valued measure (POVM) elements \cite{pkok},
\begin{align}
& \Pi_0=\sum_{k=0}^{\infty}(1-\eta)^k\left\vert k\right\rangle \left\langle k\right\vert, \nonumber \\
& \Pi_1=I-\Pi_0,
\end{align}
where $\Pi_0$ and $\Pi_1$ represent the detections of no photon or any number of photon, respectively. 
For the separated Poisson peaks in Fig. A-1, the second POVM element is effectively decomposed into
\begin{align}
 \Pi_1=\Pi^{(1)}_1+\Pi^{(2)}_1+\cdots+\Pi^{(m)}_1+\cdots ,
\end{align}
where $\Pi^{(m)}_1=\sum_{k=m_i}^{m_f}\big(1-(1-\eta)^k\big)\left\vert k\right\rangle \left\langle k\right\vert $ with the ranges 
$[m_i,m_f]$ being not overlapped, and the $m$ distinct readings of the detector corresponds the non-zero Poisson peaks of limited number. Meanwhile, the error probability of detecting nothing is
\begin{align}
P_E &=\begin{Vmatrix} \sum_{k=0}^{\infty}e^{-\left|\beta\right|^2/2} \frac{\left( \pm \beta\right)^k}{\sqrt{k!}}\left\vert k\right\rangle \Pi_0^{1/2}\left\vert \frac{\gamma e^{ik\theta}-\gamma}{\sqrt{2}}\right\rangle_{cs} \end{Vmatrix}^2 \nonumber \\
&\sim \exp{\{-2(1-e^{-\eta \gamma^2 \theta^2/2})\alpha^2 \sin^2\theta\}} \label{pe}.
\end{align}
Evidently, given the parameters $\eta \gamma^2 \theta^2 \gg 1$ and $\alpha^2 \sin^2{\theta} \gg 1$, the PND can be ideally performed with weak nonlinearity ($\theta \ll 1$).
 
The conditions on the ideal performance is different from the methods of Homodyne detection. If using $\hat{X}$-quadrature measurement, the requirement for deterministic operation is $\alpha \theta^2 \gg 1$ \cite{Barrett, munro, req1, req2}. The strength of coherent state must be much larger under the condition $\theta \ll 1$. If $\hat{P}$-quadrature measurement is used, the requirement could be improved to $\alpha \theta \gg 1$, but the operation will be probabilistic with the success probability $1/2$ \cite{req1, req2}. As a comparison, the requirement for our design is only $\alpha^2 \sin^2{\theta} \gg 1$.
Moreover, the input coherent states can be used recycled. For example, given $\theta=0.01$ and $\alpha=10^3$, the corresponding average photon number of input qubus beam is $2|\alpha|^2=2 \times 10^6$, and that of the detected coherent state is about $|\beta|^2\sim 200$. While ensuring the negligible error probability $P_E$ calculated with Eq. (\ref{pe}), such qubus beams can be used for many times, because only small portion of photons is consumed in each detection as compared with the average number of photons carried by the qubus beams themselves. With the qubus beam $\left\vert \alpha \cos^t(\theta)\right\rangle_{cs}$ and $\gamma=10^2$ after use of  $t=10^4$ times, the error probability $P_E$ is still lower than $10^{-8}$. The similar setting is also valid to the qubus beams $\left\vert \sqrt{2}\gamma \right\rangle_{cs}$ in the PND module. 

\begin{figure}[t!]
\vspace{-0cm}
\centering
\epsfig{file=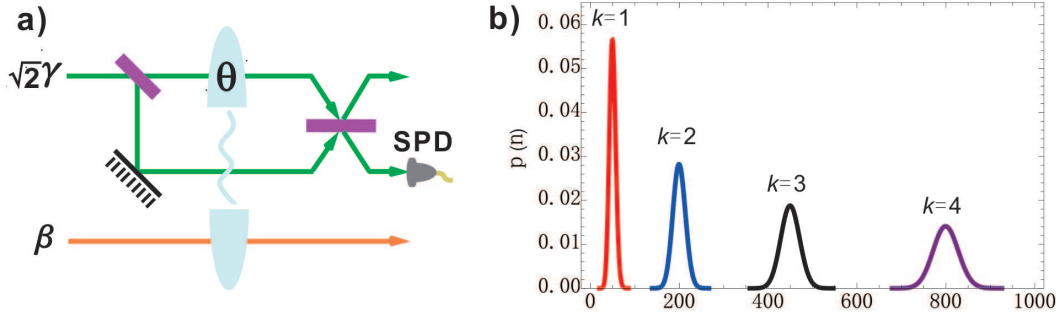,width=0.81\linewidth,clip=} 
{\vspace{-0.5cm}\caption{\label{Fig:u} a). Schematic design of the
photon number-resolving detector. The coherent state $\left\vert \beta\right\rangle_{cs}$ to be detected interacts with one of the qubus beams $\left\vert \gamma\right\rangle_{cs}\left\vert \gamma\right\rangle_{cs}$. After that, the two qubus beams interfere on a 50:50 BS. Finally,  one of the qubus beams is detected by photon number non-resolving detector.
b) The Poisson distributions of the coherent states $\left\vert \frac{\gamma e^{ik\theta}-\gamma}{\sqrt{2}}\right\rangle_{cs}$ 
for $k=1,2,3,4$. The amplitude of qubus beam is $|\gamma|=10^3$ and the cross phase shift is $\theta=0.01$. }}
\vspace{-0.3cm}
\end{figure}

\subsection*{Simplification of c-path gate}
\renewcommand{\theequation}{B-\arabic{equation}}
\renewcommand{\thefigure}{B-\arabic{figure}} 
\setcounter{equation}{0}
\setcounter{figure}{0} 
\begin{figure}[ptb]
\includegraphics[width=8.7cm]{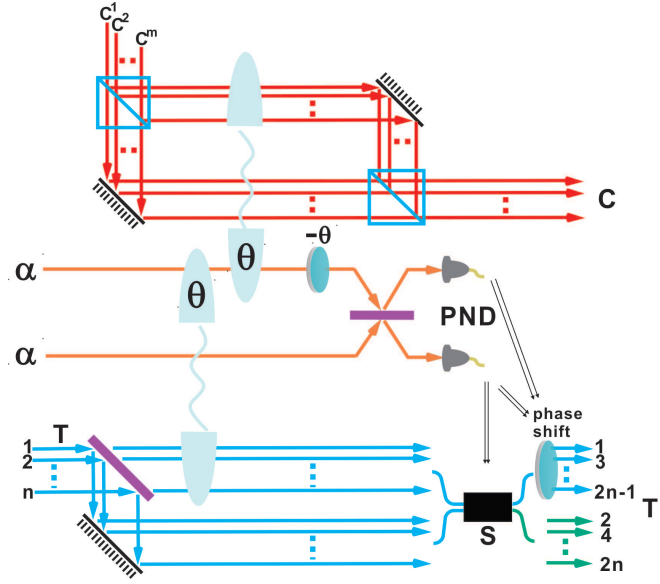}\caption{{Schematic diagram of the
modified general controlled-path gate. Compared with the original special c-path gate provided
in Fig.1, the XPM processes on the second coherent state are removed totally.
Moreover, two coherent-state components will be detected by the PNDs. In this
realization, the coherent states cannot be recycled, but only half amount of
XPM processes is necessary.}}%
\end{figure}

The realization of the general c-path gate can be simplified by removing all XPM processes
on one arm as in Fig. B-1. After the XPM processes are applied, the initial
state 
\begin{align}
\left\vert \Psi\right\rangle _{n}  &  =\left\vert H\right\rangle _{C^{1}%
,C^{2},\cdots,C^{m}}\left\vert \phi_{1}\right\rangle _{1,2,\cdots
,n}\nonumber\\
&  +\left\vert V\right\rangle _{C^{1},C^{2},\cdots,C^{m}}\left\vert \phi
_{2}\right\rangle _{1,2,\cdots,n}, \label{in} 
\end{align}
will be transformed to
\begin{align}
&  \frac{1}{\sqrt{2}}\left\{  \left\vert H\right\rangle _{C^{1},C^{2}%
,\cdots,C^{m}}\left\vert \phi_{1}\right\rangle _{1,3,\ldots,2n-1}\left\vert
\alpha e^{i\theta}\right\rangle_{cs} \left\vert \alpha\right\rangle_{cs} \right.
\nonumber\\
&  +\left\vert H\right\rangle _{C^{1},C^{2},\cdots,C^{m}}\left\vert \phi
_{1}\right\rangle _{2,4,\ldots,2n}\left\vert \alpha\right\rangle_{cs} \left\vert
\alpha\right\rangle_{cs} \nonumber\\
&  +\left\vert V\right\rangle _{C^{1},C^{2},\cdots,C^{m}}\left\vert \phi
_{2}\right\rangle _{1,3,\ldots,2n-1}\left\vert \alpha e^{i2\theta
}\right\rangle_{cs} \left\vert \alpha\right\rangle_{cs} \nonumber\\
&  \left.  +\left\vert V\right\rangle _{C^{1},C^{2},\cdots,C^{m}}\left\vert
\phi_{2}\right\rangle _{2,4,\ldots,2n}\left\vert \alpha e^{i\theta
}\right\rangle_{cs} \left\vert \alpha\right\rangle_{cs} \right\}  .
\end{align}
After a phase shifter of $-\theta$ is applied, and the two qubus beams are
interfered on a BS, the above state will be transformed to%

\begin{align}
&  \frac{1}{\sqrt{2}}\left\{  \left\vert H\right\rangle _{C^{1},C^{2}%
,\cdots,C^{m}}\left\vert \phi_{1}\right\rangle _{1,3,\ldots,2n-1}\left\vert
0\right\rangle_{cs} \left\vert \sqrt{2}\alpha\right\rangle_{cs} \right. \nonumber\\
&  +\left\vert H\right\rangle _{C^{1},C^{2},\cdots,C^{m}}\left\vert \phi
_{1}\right\rangle _{2,4,\ldots,2n}\left\vert \frac{\alpha e^{-i\theta}-\alpha
}{\sqrt{2}}\right\rangle_{cs} \left\vert \frac{\alpha e^{-i\theta}+\alpha}{\sqrt
{2}}\right\rangle_{cs} \nonumber\\
&  +\left\vert V\right\rangle _{C^{1},C^{2},\cdots,C^{m}}\left\vert \phi
_{2}\right\rangle _{1,3,\ldots,2n-1}\left\vert \frac{\alpha e^{i\theta}%
-\alpha}{\sqrt{2}}\right\rangle_{cs} \left\vert \frac{\alpha e^{i\theta}+\alpha
}{\sqrt{2}}\right\rangle_{cs} \nonumber\\
&  \left.  +\left\vert V\right\rangle _{C^{1},C^{2},\cdots,C^{m}}\left\vert
\phi_{2}\right\rangle _{2,4,\ldots,2n}\left\vert 0\right\rangle_{cs} \left\vert
\sqrt{2}\alpha\right\rangle_{cs} \right\}  ,
\end{align}
where $\left\vert \frac{\alpha e^{\mp i\theta}-\alpha}{\sqrt{2}}\right\rangle_{cs}
=\left\vert \pm i\sqrt{2}\alpha\sin\left(  \theta/2\right)  e^{\pm i\theta
/2}\right\rangle_{cs} \simeq\left\vert \pm i\alpha\theta/\sqrt{2}e^{\pm i\theta
/2}\right\rangle_{cs} ,\left\vert \frac{\alpha e^{\pm i\theta}+\alpha}{\sqrt{2}%
}\right\rangle_{cs} =\left\vert \sqrt{2}\alpha\cos\left(  \theta/2\right)  e^{\pm
i\theta/2}\right\rangle_{cs} \simeq\left\vert \sqrt{2}\alpha e^{\pm i\theta
/2}\right\rangle_{cs} .$ By the projection $\left\vert k\right\rangle \left\langle
k\right\vert $ on the first qubus beam, the target state $\left\vert
\Phi\right\rangle _{2n}$ can be obtained with the condition $k=0$. In this
case, the qubus beam could be recycled. If $k\neq0$, we will get the following
state,
\begin{align}
&  e^{-i(k\pi-k\theta)/2}\left\vert H\right\rangle _{C^{1},C^{2},\cdots,C^{m}%
}\left\vert \phi\right\rangle _{2,4,\ldots,2n}\left\vert \sqrt{2}\alpha
e^{-i\theta/2}\right\rangle_{cs} \nonumber\\
&  +e^{i(k\pi-k\theta)/2}\left\vert V\right\rangle _{C^{1},C^{2},\cdots,C^{m}%
}\left\vert \psi\right\rangle _{1,3,\ldots,2n-1}\left\vert \sqrt{2}\alpha
e^{i\theta/2}\right\rangle_{cs} .
\end{align}
Since the coherent-state component in the two terms are different, one should
measure the second qubus beam, \textit{i.e.}, the qubus beam will be lost. If
the result is $l$, the above state will be projected to%
\begin{align}
&  e^{-i(k\pi-k\theta+l\theta)/2}\left\vert H\right\rangle _{C^{1}%
,C^{2},\cdots,C^{m}}\left\vert \phi\right\rangle _{2,4,\ldots,2n}\left\vert
\sqrt{2}\alpha e^{-i\theta/2}\right\rangle_{cs} \nonumber\\
&  +e^{i(k\pi-k\theta+l\theta)/2}\left\vert V\right\rangle _{C^{1}%
,C^{2},\cdots,C^{m}}\left\vert \psi\right\rangle _{1,3,\ldots,2n-1}\left\vert
\sqrt{2}\alpha e^{i\theta/2}\right\rangle_{cs} .
\end{align}
Since the exact values of $k$ and $l$ are known, the unwanted phase factor
could be removed. Therefore the above state can be transformed to the desired
state $\left\vert \Phi\right\rangle _{2n}$ through classical feedforward.
Compared with the c-path in Fig.1, the amount of XPM processes could be
reduced to $n+m$. The cost is the qubus beam (with the probability $1/2$ when
the first detection $k\neq0$) and one more projection $\left\vert
l\right\rangle \left\langle l\right\vert $. If $n$ and $m$ is very small, one
may choose to save the coherent state. If $n$ or $m$ is large, we may choose
to reduce the amount of XPM processes.

\subsection*{The procedure of realizing the special $4$ control $1$ gate in
Fig.4}
\renewcommand{\theequation}{C-\arabic{equation}}
\renewcommand{\thefigure}{C-\arabic{figure}} 
\setcounter{equation}{0}
\setcounter{figure}{0} 

The operations shown in Fig.4 are used to realize the following 4-control-1 gate:%

\begin{equation}
F_{s,1,5}^{4}=\left(
\begin{array}
[c]{cccccc}%
\mathbb{I} &  &  &  &  & \\
& \vdots &  &  &  & \\
&  & \mathbb{I} &  &  & \\
&  &  & U_{13} &  & \\
&  &  &  & \vdots & \\
&  &  &  &  & U_{16}%
\end{array}
\right)  .
\end{equation}
Obviously, the single qubit operations $U_{13},\ldots,U_{16}$ will be
implemented to the target photon only when the first two photons are all in
the state $\left\vert V\right\rangle $. The initial state can be described as
follows:
\begin{align}
&  \left\vert HHHH\right\rangle \left\vert \phi_{1}\right\rangle +\left\vert
HHHV\right\rangle \left\vert \phi_{2}\right\rangle +\left\vert
HHVH\right\rangle \left\vert \phi_{3}\right\rangle \nonumber\\
&  +\left\vert HHVV\right\rangle \left\vert \phi_{4}\right\rangle +\left\vert
HVHH\right\rangle \left\vert \phi_{5}\right\rangle +\left\vert
HVHV\right\rangle \left\vert \phi_{6}\right\rangle \nonumber\\
&  +\left\vert HVVH\right\rangle \left\vert \phi_{7}\right\rangle +\left\vert
HVVV\right\rangle \left\vert \phi_{8}\right\rangle +\left\vert
VHHH\right\rangle \left\vert \phi_{9}\right\rangle \nonumber\\
&  +\left\vert VHHV\right\rangle \left\vert \phi_{10}\right\rangle +\left\vert
VHVH\right\rangle \left\vert \phi_{11}\right\rangle +\left\vert
VHVV\right\rangle \left\vert \phi_{12}\right\rangle \nonumber\\
&  +\left\vert VVHH\right\rangle \left\vert \phi_{13}\right\rangle +\left\vert
VVHV\right\rangle \left\vert \phi_{14}\right\rangle +\left\vert
VVVH\right\rangle \left\vert \phi_{15}\right\rangle \nonumber\\
&  +\left\vert VVVV\right\rangle \left\vert \phi_{16}\right\rangle ,
\end{align}
where $\left\vert \phi_{i}\right\rangle =\alpha_{i}\left\vert H\right\rangle
+\beta_{i}\left\vert V\right\rangle $, $\sum\limits_{i=1}^{16}\left(
\left\vert \alpha_{i}\right\vert ^{2}+\left\vert \beta_{i}\right\vert
^{2}\right)  =1$. First, let the first photon controls the second photon
through the first c-path gate. After that, let the first spatial mode $1$ of
the second photon passed through a PBS and perform a $\sigma_{x}$ operation on
the $1^{\prime}$ mode, yielding the following state
\begin{align}
&  \left(  \left\vert HHHH\right\rangle \left\vert \phi_{1}\right\rangle
+\left\vert HHHV\right\rangle \left\vert \phi_{2}\right\rangle +\left\vert
HHVH\right\rangle \left\vert \phi_{3}\right\rangle \right. \nonumber\\
&  \left.  +\left\vert HHVV\right\rangle \left\vert \phi_{4}\right\rangle
\right)  _{1}+\left(  \left\vert HHHH\right\rangle \left\vert \phi
_{5}\right\rangle +\left\vert HHHV\right\rangle \left\vert \phi_{6}%
\right\rangle \right. \nonumber\\
&  \left.  +\left\vert HHVH\right\rangle \left\vert \phi_{7}\right\rangle
+\left\vert HHVV\right\rangle \left\vert \phi_{8}\right\rangle \right)
_{1^{\prime}}+\left(  \left\vert VHHH\right\rangle \left\vert \phi
_{9}\right\rangle \right. \nonumber\\
&  +\left\vert VHHV\right\rangle \left\vert \phi_{10}\right\rangle +\left\vert
VHVH\right\rangle \left\vert \phi_{11}\right\rangle +\left\vert
VHVV\right\rangle \left\vert \phi_{12}\right\rangle \nonumber\\
&  +\left\vert VVHH\right\rangle \left\vert \phi_{13}\right\rangle +\left\vert
VVHV\right\rangle \left\vert \phi_{14}\right\rangle +\left\vert
VVVH\right\rangle \left\vert \phi_{15}\right\rangle \nonumber\\
&  \left.  +\left\vert VVVV\right\rangle \left\vert \phi_{16}\right\rangle
\right)  _{2},
\end{align}
where the subscripts outside the bracket denote the spatial modes of the
second photon. Second, using the three spatial modes of the second photon to
control the photon $3,4,5$ by three c-path gates, one will obtain the
following state%

\begin{align}
&  \left(  \left\vert HHHH\right\rangle \left\vert \phi_{1}\right\rangle
+\left\vert HHHV\right\rangle \left\vert \phi_{2}\right\rangle +\left\vert
HHVH\right\rangle \left\vert \phi_{3}\right\rangle \right. \nonumber\\
&  \left.  +\left\vert HHVV\right\rangle \left\vert \phi_{4}\right\rangle
\right)  _{1111}+\left(  \left\vert HHHH\right\rangle \left\vert \phi
_{5}\right\rangle +\left\vert HHHV\right\rangle \left\vert \phi_{6}%
\right\rangle \right. \nonumber\\
&  \left.  +\left\vert HHVH\right\rangle \left\vert \phi_{7}\right\rangle
+\left\vert HHVV\right\rangle \left\vert \phi_{8}\right\rangle \right)
_{1^{\prime}111}+\left(  \left\vert VHHH\right\rangle \left\vert \phi
_{9}\right\rangle \right. \nonumber\\
&  \left.  +\left\vert VHHV\right\rangle \left\vert \phi_{10}\right\rangle
+\left\vert VHVH\right\rangle \left\vert \phi_{11}\right\rangle +\left\vert
VHVV\right\rangle \left\vert \phi_{12}\right\rangle \right)  _{2111}%
\nonumber\\
&  +\left(  \left\vert VVHH\right\rangle \left\vert \phi_{13}\right\rangle
+\left\vert VVHV\right\rangle \left\vert \phi_{14}\right\rangle +\left\vert
VVVH\right\rangle \left\vert \phi_{15}\right\rangle \right. \nonumber\\
&  \left.  +\left\vert VVVV\right\rangle \left\vert \phi_{16}\right\rangle
\right)  _{2222},
\end{align}
where the subscripts outside the bracket denote the spatial modes of the
photons except for the first one. Obviously, the photons $3,4,5$ will be
separated into the spatial mode $2$ only when the first two photons are all in
the state $\left\vert V\right\rangle $. Therefore, following the processes in
the part of "general $(n-1)$-control-1 unitary operation" to operate on all the spatial modes $2$ of the photons $3,4,5$,
one will obtain the state
\begin{align}
&  \left(  \left\vert HHHH\right\rangle \left\vert \phi_{1}\right\rangle
+\left\vert HHHV\right\rangle \left\vert \phi_{2}\right\rangle +\left\vert
HHVH\right\rangle \left\vert \phi_{3}\right\rangle \right. \nonumber\\
&  \left.  +\left\vert HHVV\right\rangle \left\vert \phi_{4}\right\rangle
\right)  _{1111}+\left(  \left\vert HHHH\right\rangle \left\vert \phi
_{5}\right\rangle +\left\vert HHHV\right\rangle \left\vert \phi_{6}%
\right\rangle \right. \nonumber\\
&  \left.  +\left\vert HHVH\right\rangle \left\vert \phi_{7}\right\rangle
+\left\vert HHVV\right\rangle \left\vert \phi_{8}\right\rangle \right)
_{1^{\prime}111}+\left(  \left\vert VHHH\right\rangle \left\vert \phi
_{9}\right\rangle \right. \nonumber\\
&  \left.  +\left\vert VHHV\right\rangle \left\vert \phi_{10}\right\rangle
+\left\vert VHVH\right\rangle \left\vert \phi_{11}\right\rangle +\left\vert
VHVV\right\rangle \left\vert \phi_{12}\right\rangle \right)  _{2111}%
\nonumber\\
&  +\left(  \left\vert VVHH\right\rangle \otimes U_{13}\left\vert \phi
_{13}\right\rangle +\left\vert VVHV\right\rangle \otimes U_{14}\left\vert
\phi_{14}\right\rangle \right. \nonumber\\
&  \left.  +\left\vert VVVH\right\rangle \otimes U_{15}\left\vert \phi
_{15}\right\rangle +\left\vert VVVV\right\rangle \otimes U_{16}\left\vert
\phi_{16}\right\rangle \right)  _{2222}.
\end{align}
Finally, the inverse merging gates are applied to the corresponding single
photons to transform the above state to the target state
\begin{align}
&  \left\vert HHHH\right\rangle \left\vert \phi_{1}\right\rangle +\left\vert
HHHV\right\rangle \left\vert \phi_{2}\right\rangle +\left\vert
HHVH\right\rangle \left\vert \phi_{3}\right\rangle \nonumber\\
&  +\left\vert HHVV\right\rangle \left\vert \phi_{4}\right\rangle +\left\vert
HHHH\right\rangle \left\vert \phi_{5}\right\rangle +\left\vert
HHHV\right\rangle \left\vert \phi_{6}\right\rangle \nonumber\\
&  +\left\vert HHVH\right\rangle \left\vert \phi_{7}\right\rangle +\left\vert
HHVV\right\rangle \left\vert \phi_{8}\right\rangle +\left\vert
VHHH\right\rangle \left\vert \phi_{9}\right\rangle \nonumber\\
&  +\left\vert VHHV\right\rangle \left\vert \phi_{10}\right\rangle +\left\vert
VHVH\right\rangle \left\vert \phi_{11}\right\rangle +\left\vert
VHVV\right\rangle \left\vert \phi_{12}\right\rangle \nonumber\\
&  +\left\vert VVHH\right\rangle \otimes U_{13}\left\vert \phi_{13}%
\right\rangle +\left\vert VVHV\right\rangle \otimes U_{14}\left\vert \phi
_{14}\right\rangle \nonumber\\
&  +\left\vert VVVH\right\rangle \otimes U_{15}\left\vert \phi_{15}%
\right\rangle +\left\vert VVVV\right\rangle \otimes U_{16}\left\vert \phi
_{16}\right\rangle .
\end{align}
This is how a special 4-control-1 gate is implemented.